\documentclass[pra,twocolumn,aps,groupedaddress,longbibliography,superscriptaddress]{revtex4-2}

\usepackage{graphicx}

\usepackage{subfigure}

\usepackage{float}
\usepackage{hyperref}
\hypersetup{
      colorlinks=true,
      citecolor=blue,
      linkcolor=blue,
      urlcolor=blue}
\usepackage{amsmath}
\usepackage{amsfonts}
\usepackage{amssymb}
\usepackage{braket}
\usepackage{bm}
\usepackage{times}
\usepackage{amsmath}
\usepackage{braket}
\usepackage{ragged2e}
\usepackage{graphicx}
\usepackage{xcolor}
\usepackage{nicefrac} 
\usepackage{lipsum}
\usepackage{nameref}
\usepackage{hyperref}
\usepackage{textcomp} 
\usepackage{urwchancal}
\usepackage{upgreek} 
\usepackage{amsmath} 
\usepackage{here}
\usepackage{siunitx}
\usepackage{booktabs}
\usepackage{longtable}
\usepackage{scalerel}
\usepackage{tikz}
\usetikzlibrary{svg.path}
\usepackage{nicefrac} 
\usepackage{upgreek}
\usepackage{algpseudocode}
\usepackage{algorithm}

\graphicspath{{figs/}}

\DeclareSIUnit\gauss{G}

\definecolor{orcidlogocol}{HTML}{A6CE39}
\tikzset{
  orcidlogo/.pic={
    \fill[orcidlogocol] svg{M256,128c0,70.7-57.3,128-128,128C57.3,256,0,198.7,0,128C0,57.3,57.3,0,128,0C198.7,0,256,57.3,256,128z};
    \fill[white] svg{M86.3,186.2H70.9V79.1h15.4v48.4V186.2z}
                 svg{M108.9,79.1h41.6c39.6,0,57,28.3,57,53.6c0,27.5-21.5,53.6-56.8,53.6h-41.8V79.1z M124.3,172.4h24.5c34.9,0,42.9-26.5,42.9-39.7c0-21.5-13.7-39.7-43.7-39.7h-23.7V172.4z}
                 svg{M88.7,56.8c0,5.5-4.5,10.1-10.1,10.1c-5.6,0-10.1-4.6-10.1-10.1c0-5.6,4.5-10.1,10.1-10.1C84.2,46.7,88.7,51.3,88.7,56.8z};
  }
}

\newcommand\orcidicon[1]{\href{https://orcid.org/#1}{\mbox{\scalerel*{
\begin{tikzpicture}[yscale=-1,transform shape]
\pic{orcidlogo};
\end{tikzpicture}
}{|}}}}

\begin{document}

\title{Optical Manipulation of Spin States in Ultracold Magnetic Atoms via an Inner-Shell Hz Transition}

\author{F. Claude\,\orcidicon{0000-0003-1118-7379}}
\affiliation{Institut f\"{u}r Quantenoptik und Quanteninformation, \"Osterreichische Akademie der \\ Wissenschaften, 6020 Innsbruck, Austria}

\author{L. Lafforgue\,\orcidicon{0009-0000-8234-6328}}
\affiliation{Universit\"at Innsbruck, Institut f\"{u}r Experimentalphysik, 6020 Innsbruck, Austria}

\author{J. J. A. Houwman\,\orcidicon{0009-0003-3342-2322}}
\affiliation{Universit\"at Innsbruck, Institut f\"{u}r Experimentalphysik, 6020 Innsbruck, Austria}

\author{M. J. Mark\,\orcidicon{0000-0001-8157-4716}}
\affiliation{Institut f\"{u}r Quantenoptik und Quanteninformation, \"Osterreichische Akademie der \\ Wissenschaften, 6020 Innsbruck, Austria}
\affiliation{Universit\"at Innsbruck, Institut f\"{u}r Experimentalphysik, 6020 Innsbruck, Austria}

\author{F. Ferlaino\,\orcidicon{0000-0002-3020-6291}}
\affiliation{Institut f\"{u}r Quantenoptik und Quanteninformation, \"Osterreichische Akademie der \\ Wissenschaften, 6020 Innsbruck, Austria}
\affiliation{Universit\"at Innsbruck, Institut f\"{u}r Experimentalphysik, 6020 Innsbruck, Austria}

\date{\today}

\begin{abstract}
Lanthanides, like erbium and dysprosium, have emerged as powerful platforms for quantum-gas research due to their diverse properties, including a significant large spin manifold in their absolute ground state. However, effectively exploiting the spin richness necessitates precise manipulation of spin populations, a challenge yet to be fully addressed in this class of atomic species. In this work, we present an all-optical method for deterministically controlling the spin composition of a dipolar bosonic erbium gas, based on a clock-like transition in the telecom window at \SI{1299}{\nano \meter}. The atoms can be prepared in just a few tens of microseconds in any spin-state composition using a sequence of Rabi-pulse pairs, selectively coupling Zeeman sublevels of the ground state with those of the long-lived clock-like state. Finally, we demonstrate that this transition can also be used to create spin-selective light shifts, thus fully suppressing spin-exchange collisions. These experimental results unlock exciting possibilities for implementing advanced spin models in isolated, clean and fully controllable lattice systems.
\end{abstract}

\maketitle

\section{Introduction}

Models of interacting spins are iconic in condensed matter physics, offering a rich framework for probing fundamental phenomena in magnetic materials, such as quantum magnetism~\cite{Sachdev2008qma}, exotic topological phases~\cite{Hasan2010cti, Qi2011tia}, quantum criticality~\cite{Sachdev2011qpt}, and spin glasses~\cite{Binder1986sge}. They are investigated in systems of many correlated  quantum particles, e.\,g.\,arrays of Rydberg atoms~\cite{Saffman2010qiw}, polar molecules~\cite{Bohn2017cmp}, trapped ions~\cite{Blatt2012qsw}, magnetic atoms~\cite{Griesmaier2005bec,Lu2011sdb,Aikawa2012bec} and coupled microcavities~\cite{Hartmann2008qmb, Ji2007qmd}. In the case of alkali neutral atoms, the spin dynamics are mediated by super-exchange interactions, which are typically weak compared to the other energy scales in the systems~\cite{Trotzky2008tro}. Differently, atoms with a large magnetic moment, such as dysprosium~\cite{Burdick2015fso,Chalopin2018qes,Bouhiron2024raq}, erbium~\cite{Baier2018roa,Patscheider2020cde}, and chromium~\cite{cr1_dePaz2013nqm,cr2_dePaz2016psd,cr3_Lepoutre2019oeq}, enable strong interactions through long-range and anisotropic dipole-dipole forces. This feature, combined with the large spin-manifold possessed especially by lanthanides, provides access to an unprecedented class of long-range-interacting spin-lattice models~\cite{Norcia2021dia, Chomaz2022dpa, Dutta2015nsh}.
%StamperKurn2013sbg

These outstanding properties in lanthanides also come with some challenging aspects. Firstly, the strong magnetic dipole-dipole interaction (DDI) leads to typically fast spin-spin collisions, encompassing both spin-conserving two-body exchange processes and spin-non-conserving relaxation processes~\cite{Pasquiou2010cod,Burdick2015fso, Barral2023ctd}. 
Secondly, the combination of a large DDI and the orbital anisotropy of van-der-Waals interactions, inherent to non-$S$ ground-state atoms~\cite{Kotochigova2014cib}, leads to an extraordinarily dense spectrum of Fano-Feshbach resonances between spin-polarized atoms in their absolute ground state~\cite{Frisch2014qci, Baumann2014ool, Maier2015eoc, Khlebnikov2019rtc} and even more so in spin mixtures~\cite{Baier2018roa}.
Finally, in bosonic lanthanides, the absence of hyperfine structure — due to the zero nuclear spin — removes any quadratic term in the Zeeman splitting, prohibiting spin selectivity in standard radio-frequency (RF) protocols for spin-state initialization~\cite{cr2_dePaz2016psd, cr3_Lepoutre2019oeq}. While the first and second aspect can be largely mitigated by the use of an optical lattice~\cite{Baier2018roa,Patscheider2020cde, Pasquiou2010cod}, the third requires the development of alternative approaches based on optical transitions. 

In the present work, we demonstrate a novel approach to deterministically initialize atoms in any spin configuration using an inner-shell clock-like transition~\cite{Patscheider2021ooa}. In Sec.~\ref{sec2}, we introduce our spin manipulation protocol, which relies on the resonant and selective coupling of Zeeman sublevels in the ground state to those in the long-lived clock-like state with an ultra-narrow laser. In Sec.~\ref{sec3}, we demonstrate the versatility of our method through the generation and observation of various spin mixtures within the 13 internal states of bosonic erbium. In Sec.~\ref{sec4}, we achieve selective manipulation of the energy of Zeeman sublevels via a light shift induced by the ultra-narrow laser. This method effectively suppresses spin-exchange processes in spinor Bose-Einstein condensates (BECs) of erbium.

\section{Experimental setup}\label{sec2}

Our experimental sequence starts with the generation of an ultracold gas of bosonic $^{166}$Er atoms~\cite{Aikawa2012bec}. The particles are spin-polarized in the lowest Zeeman sublevel of the ground state $\ket{J=6, m_{J}=-6}$. Here, $J$ is the angular momentum quantum number and $m_J$ its projection along the quantization axis. After loading a narrow-line magneto-optical trap~\cite{Frisch2012nlm}, we transfer the atoms into a crossed optical dipole trap and perform further cooling via standard evaporation steps. A homogeneous magnetic field $B$ of \SI{1.9}{\gauss} is applied along $z$ during the evaporation to preserve spin polarization. We end up with a thermal cloud of about $4 \times 10^{4}$ atoms at a temperature of \SI{197(2)}{\nano\kelvin}, which lies just above the critical temperature for Bose-Einstein condensation in our final trap with frequencies ($\omega_{x}$, $\omega_{y}$, $\omega_{z}$) = 2$\pi \times \left[122(2), 89(2), 167(1) \right]$\SI{}{\hertz} (gravity is pointing along $z$).

\begin{figure}
         \includegraphics[width=1\columnwidth]{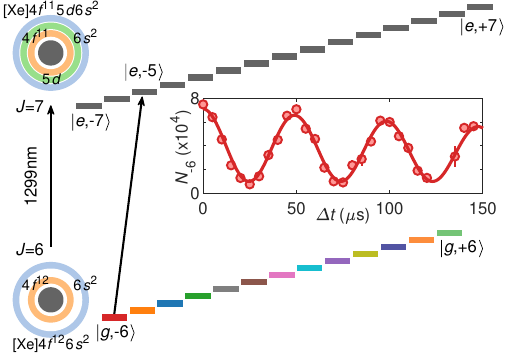}
        \caption{ Illustration of the manifolds of Zeeman sublevels of the bosonic Er ground state $\left(\ket{J=6, m_J}\right)$ and the clock-like state at \SI{1299}{\nano \meter} $\left(\ket{J'=7, m_{J'}}\right)$ and their electronic configuration. Inset: Coherent Rabi oscillation of the transition $\ket{g, -6}\leftrightarrow\ket{e, -5}$. The atom number in $\ket{g, -6}$ is plotted as a function of the clock laser illumination time $\Delta t$. We fit a damped sinusoid to the data (solid curve), giving a Rabi frequency of $\Omega_R/2\pi=$\SI{20.3(1)}{\kilo\hertz} and a damping time of $\tau=$\SI{525(118)}{\micro \second}. Later for our transfer scheme, we will use illumination pulses of $\Delta t\leq \pi/\Omega_R \simeq$ \SI{25}{\micro \second}.}
    \label{fig:fig1}
\end{figure}

Bosonic erbium has zero nuclear spin ($I = 0$) and thus it possesses no hyperfine structure. As a consequence, the energy splitting between any pair of consecutive levels among the $2J + 1 = 13$ spin states available in the ground state is identical. A standard radio-frequency sweep or pulse would then couple all spin levels, preventing selective initialization of  specific spin states. To overcome this limitation, we have developed a new approach based on a sequence of coherent Rabi pulses operating on the ultra-narrow optical transition at \SI{1299}{\nano\meter}. The 1299-nm transition has an inner-shell orbital character since it corresponds to the promotion of one $4f$ electron from the $[\mathrm{Xe}] 4f^{12}6s^{2}$ state into the $5d$ shell of the $[\mathrm{Xe}] 4f^{11}5d6s^{2}$ state. Due to its very narrow linewidth $\Gamma = 2 \pi \times$\SI{0.9(1)}{\hertz} \cite{Patscheider2021ooa}, it is therefore referred to as a clock-like transition. To distinguish between the Zeeman sublevels in the ground- and excited-state manifold, we will denote them as $\ket{g, m_J}$ and $\ket{e, m_{J'}}$, respectively. 

Figure\,\ref{fig:fig1} depicts the associated Zeeman manifolds of the system. The difference in Landé g-factor between the ground and clock-like state results in a differential Zeeman splitting of $\left(g_{J'}-g_J \right)\mu_B$ = \SI{134}{\kilo\hertz\per\gauss} \footnote{We measured experimentally that $g_J$ and $g_{J'}$ are equal to $1.163801(1)$ and $1.2599(5)$ respectively.}, where $\mu_B$ is the Bohr magneton. Since the splitting is large compared to the natural linewidth of the clock-like state ($\approx 24\Gamma /$\SI{}{\milli \gauss}), each transition can be individually addressed, even at magnetic fields below \SI{1}{\milli \gauss}. An ultra-narrow laser with a linewidth of \SI{10}{\hertz} drives the clock-like transition. It is stabilized with a high-finesse, environmentally isolated cavity featuring an Allan deviation of \SI{3.1e-15}{\per \second} and a drift rate below \SI{15.5}{\milli\hertz\per\second}\cite{Patscheider2021ooa}. The laser light contains contributions from all polarizations, such that $\sigma^+$, $\sigma^-$, and $\pi$ transitions can be addressed. As an example, the inset in Fig.\,\ref{fig:fig1} shows the Rabi oscillation corresponding to the $\sigma^+$ transition $\ket{g, -6}\leftrightarrow\ket{e, -5}$.

\section{Coherent control of spin states}\label{sec3}

To manipulate the spin composition of our dipolar gas, we design a protocol based on a sequence of consecutive pairs of coherent Rabi pulses. Each pair of pulses transfers atoms between Zeeman sublevels in the ground state \textit{via} an intermediate clock-like excited state, $\ket{g, m_J}\to\ket{e, m_{J'}}\to\ket{g, m_J+1}$ or $\ket{g, m_J+2}$, enabling an ascent through the ladder of levels in the ground-state manifold.

\begin{figure}
        \includegraphics[width=1.0\columnwidth]{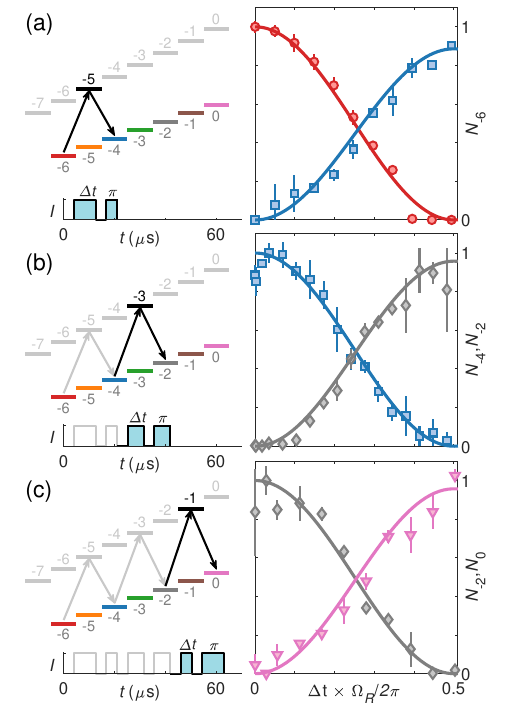}
        \caption{Spin manipulation protocol. (a-c) Left panels: 1299-nm pulse scheme for transferring atoms from $\ket{g, -6}$ to $\ket{g, 0}$. Each pair of pulses increments $m_J$ by +2 in the ground state. The first pair transfers the atoms from $\ket{g, -6}$ to $\ket{g, -4}$ (a), the second from $\ket{g, -4}$ to $\ket{g, -2}$ (b) and the third from $\ket{g, -2}$ to $\ket{g, 0}$ (c). Right panels: Measured spin populations in states $\ket{g, -6}$ (circles), $\ket{g, -4}$ (squares), $\ket{g, -2}$ (diamonds) and $\ket{g, 0}$ (hexagons) as a function of the duration $\Delta t$ of the first (a), third (b) and fifth (c) pulse of the sequence, normalized by the Rabi frequency $\Omega_R$ of the related transition. Data points consist of a minimum of three individual realizations and are normalized by the atom numbers at $\Delta t =$ \SI{0}{}.}
    \label{fig:fig2}
\end{figure}

The duration of the first pulse of each Rabi pair controls the amount of population to be transferred, whereas the second pulse brings the whole fraction of atoms in $\ket{e, m_{J'}}$ into the desired spin level of the ground state manifold $\ket{g, m_{J}}$ ($\pi$-pulse). Therefore, by precisely adjusting the duration of the first pulse to within half of a Rabi oscillation (${\Delta t \Omega_R / 2 \pi < 0.5}$), we can transfer any fraction of atoms to the excited Zeeman sublevels. Finally, we probe the atom numbers through absorption imaging combined with a standard Stern-Gerlach procedure; see Appendix\,\ref{appA}.

\begin{figure}
         \includegraphics[width=1.\columnwidth]{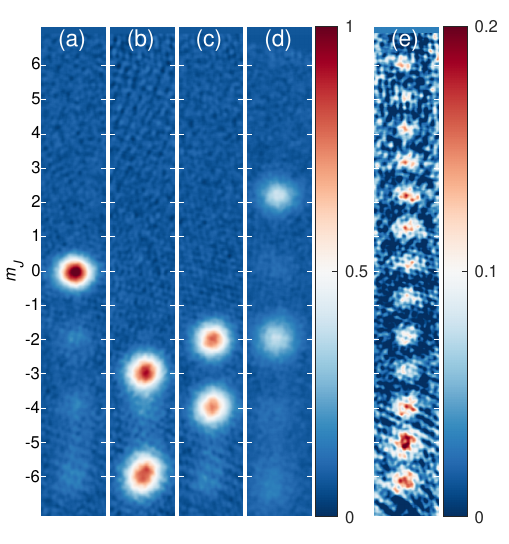}
        \caption{Stern-Gerlach absorption images of $^{166}$Er spin states. (a-d) Preparations of $\ket{g, 0}$, $\ket{g, -6} + \ket{g, -3}$, $\ket{g, -4} + \ket{g, -2}$ and $\ket{g, -2} + \ket{g, +2}$ states, obtained with different sequences of 1299-nm pulses. (e) Preparation obtained with a resonant RF-pulse. The atomic densities indicated by the colored bars are normalized by the $\ket{g, 0}$ peak density in panel (a). To separate the spin populations, we apply a magnetic field gradient of \SI{20}{\gauss\per\centi\meter} for the first \SI{5.5}{\milli\second} of time of flight.}
    \label{fig:fig3}
\end{figure}

Figure\,\ref{fig:fig2} illustrates an example of our protocol, wherein atoms in the lowest spin state $\ket{g, -6}$ are transferred to the $\ket{g, 0}$ state via three pairs of Rabi pulses. Specifically, we start by finely tuning the duration of the first Rabi pulse $\Delta t$ to transfer the desired amount of population to $\ket{e, -5}$. Subsequently, we apply a $\pi$-pulse to transfer all the atoms from $\ket{e, -5}$ to $\ket{g, -4}$ [Fig.\,\ref{fig:fig2}(a)]. We then repeat the same sequence two times, such that atoms are incrementally transferred to $\ket{g, -2}$ [Fig.\,\ref{fig:fig2}(b)] and $\ket{g, 0}$ [Fig.\,\ref{fig:fig2}(c)]. For a full transfer of all atoms to $\ket{g,0}$, we use only pairs of $\pi$-pulses and record an overall efficiency of \SI{85}{\percent}.
Figure\,\ref{fig:fig3} shows examples of absorption images of different spin compositions: all atoms in the $\ket{g, 0}$ state (a), balanced spin mixtures $\ket{g, -6} + \ket{g, -3}$ (b), $\ket{g, -4} + \ket{g, -2}$ (c), and $\ket{g, -2} + \ket{g, +2}$ (d). For comparison, Fig.\,\ref{fig:fig3}(e) shows an image of the spin composition obtained after applying a RF-pulse, highlighting the simultaneous coupling of all sublevels in the ground-state manifold.

\section{INHIBITION OF SPIN-EXCHANGE}\label{sec4}

Further control of spin dynamics can be accomplished by inducing a light-shift of the atomic energy levels with a laser~\cite{Anderson2000vpi}. Our clock-like transition not only enables the deterministic preparation of any spin composition, as depicted in Fig.\,\ref{fig:fig3}, but also facilitates precise adjustment of the energy of each individual Zeeman sublevel. This is achieved through a spin-selective light shift induced by the 1299-nm light. As a proof of concept, we demonstrate our ability to close a spin-exchange depolarization channel by inducing a light shift to a specific Zeeman sublevel.

Magnetic atoms in higher spin states undergo collisional losses mainly driven by two mechanisms~\cite{Chomaz2022dpa}: spin-exchange and spin relaxation. The first process conserves the total magnetization and energy while the spin states of the atoms change their projection quantum numbers as $\Delta m_J=\pm1$. Therefore, if the energy splitting between consecutive Zeeman sublevels is equal, as in the case of bosonic erbium, spin-exchange processes lead to a depolarization. Spin relaxation processes on the other hand do not preserve the total magnetization and energy. Instead, these processes correspond to one or both atoms lowering their spin projection quantum number $\Delta m_J=-1$. The Zeeman energy difference is released as kinetic energy, typically large enough to expel both atoms from the trap.

\begin{figure}
         \includegraphics[width=1.\columnwidth]{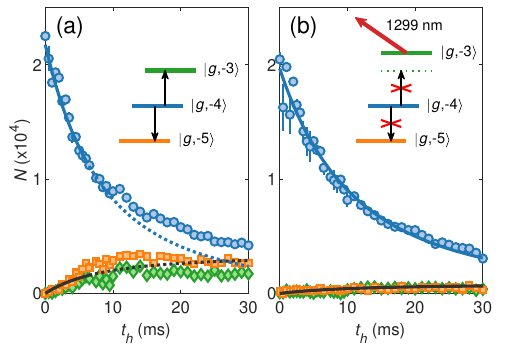}
        \caption{Evolution of the atom numbers in $\ket{g,-5}$ (squares), $\ket{g,-4}$ (circles), and $\ket{g,-3}$ (diamonds) states with (a) and without (b) spin-exchange mechanisms. The solid lines show the results of our rate-equation model fitted to the data within $t_h < 7\,$ms (a) and $t_h < 30\,$ms (b). The dotted lines in (a) correspond to the results of our rate-equation model outside the fitting range.
        Without a light shift (a), we extract the relaxation rate ${\beta_{\textrm{sr}} = \SI{3.8(1)e-13}{\cubic \centi \meter\per \second}}$ and the exchange rate ${\beta_{\textrm{ex}} = \SI{1.5(1)e-13}{\cubic \centi \meter \per \second}}$. With light shift (b), ${\beta_{\textrm{sr}} = \SI{3.9(2)e-13}{\cubic \centi \meter \per\second}}$ and ${\beta_{\textrm{ex}} = \SI{0.3(2)e-13}{\cubic \centi \meter \per\second}}$. 
        }
    \label{fig:fig4}
\end{figure}

To study these collision processes, we prepare a BEC in $\ket{g,-4}$ and follow the time evolution of the atom number in the dipole trap by varying the hold time $t_h$. As shown in Fig.\,\ref{fig:fig4}(a), we observe a fast decay of atoms in $\ket{g,-4}$, and simultaneously an increase in population in the neighboring states $\ket{g,-5}$ and $\ket{g,-3}$, which saturates for longer hold times.
Such an increase is a direct consequence of spin-exchange processes.

We quantify the decay and growth rates of atoms in the different spin-states with a simple rate-equation model. We account only for two-body processes, assuming that one-body losses due to background collisions and three-body losses are much slower; see Appendix\,\ref{appB}. Additionally, we only consider the early time dynamics in the trap ($t_h <$ \SI{7}{\milli \second}), during which the populations in $\ket{g,-5}$ and $\ket{g,-3}$ are smaller compared to that in $\ket{g,-4}$, allowing us to disregard secondary two-body losses originating from these states. Integrating over the condensate volume, the spin populations evolve according to \cite{Satoshi2009sdi}
\begin{align}
\dot{N}_{-4} = -\alpha \left( \beta_{\textrm{sr}} + \beta_{\textrm{ex}} \right) N_{-4}^{7/5}, \label{2b-population-4}\\
\dot{N}_{-5} = \dot{N}_{-3} = \alpha \dfrac{\beta_{\textrm{ex}}}{2} N_{-4}^{7/5}, \label{2b-population-35}
\end{align}
where $\alpha = \nicefrac{15^{\frac{2}{5}}}{14 \pi} \left( \nicefrac{m \bar{\omega}}{\hbar \sqrt{a_{-6}}} \right)^{\frac{6}{5}}$. Here, $\bar{\omega}$ is the average trap frequency and $a_{-6}$ is the scattering length in $\ket{g,-6}$. The parameters $\beta_{\textrm{sr}}$ and $\beta_{\textrm{ex}}$ represent the spin relaxation and exchange rates, respectively.

Fitting the atom number in Fig.\,\ref{fig:fig4}(a) to Eq.\,\eqref{2b-population-4} and \eqref{2b-population-35}, we find good agreement for the early time dynamics. From this analysis, we extract spin relaxation and exchange rates: ${\beta_{\textrm{sr}} = \SI{3.8(1)e-13}{\cubic \centi \meter\per \second}}$ and ${\beta_{\textrm{ex}} = \SI{1.5(1)e-13}{\cubic \centi \meter \per \second}}$, respectively. However, for longer trapping times, our model fails to accurately reproduce the experimental data due to collisions between atoms in different Zeeman sublevels.

Since spin-exchange is a resonant process, requiring equal Zeeman splitting between the starting and the two final neighboring states, it can be inhibited by lifting the degeneracy. This can be engineered by inducing a light shift to $\ket{g,-3}$, as illustrated in the inset of Fig.\,\ref{fig:fig4}(b). 
In our experiment, we create a light shift of about \SI{3}{\kilo \hertz} with the 1299-nm light tuned on resonance to the transition $\ket{g, -3}\leftrightarrow\ket{e, -4}$ by continuously illuminating the sample during the hold time with a beam of typically \SI{10}{\micro \watt}. 

As illustrated in Fig.\,\ref{fig:fig4}(b), with this protocol, we can effectively eliminate spin-exchange collisions, keeping the population in the two neighboring states negligible.  The fitting of atom numbers to Eq.\,\eqref{2b-population-4} and \eqref{2b-population-35} shows excellent agreement with the experimental data across the entire measured time range. We extract relaxation and exchange rates: ${\beta_{\textrm{sr}} = \SI{3.9(2)e-13}{\cubic \centi \meter \per\second}}$ and ${\beta_{\textrm{ex}} = \SI{0.3(2)e-13}{\cubic \centi \meter \per\second}}$, respectively, confirming the successful inhibition of spin-exchange processes. 

\section{Conclusion}
 
In conclusion, we demonstrate unprecedented control over the internal spin states of erbium. 
Our protocol uses pairs of coherent Rabi pulses, transferring population from one spin state to another via a clock-like transition at \SI{1299}{\nano \meter}.
Moreover, we show that by driving this transition we can selectively induce light shifts of a desired Zeeman sublevel. This enables us to fully close the loss channels given by spin-exchange processes, thereby preserving the polarization of our samples. 
By combining this technique with our ability to prepare atoms in any spin state, we can now flexibly adjust the spin size of our system.
While the losses induced by spin relaxation remain a limitation in bulk bosonic systems, they can be mitigated by confining atoms to thin layers \cite{Barral2023ctd} or optical lattices \cite{Pasquiou2010cod}. 

As our protocol is independent of the magnitude of the magnetic field, it is especially appealing for atoms exhibiting a dense Feshbach spectrum, such as the lanthanides. Moreover, our preparation scheme can be directly generalized to dysprosium, which has a similar orbital clock-like transition at \SI{1001}{\nano \meter}~\cite{Petersen2020sot}, and ytterbium, possessing one at \SI{431}{\nano \meter}~\cite{Ishiyama2023ooa}.

Therefore, our findings provide a solid groundwork for forthcoming experimental studies of diverse dipolar spinor gases \cite{StamperKurn2013sbg} and lattice spin models \cite{Trefzger2011udg}.

\section{ACKNOWLEDGMENTS}

We thank Thomas Bland for discussions and careful reading of the manuscript. We acknowledge support from the European Research Council through the Advanced Grant DyMETEr (\href{https://doi.org/10.3030/101054500}{10.3030/101054500}), a NextGeneration EU Grant AQuSIM through the Austrian Research Promotion Agency (FFG) (No.\,FO999896041), and the Austrian Science Fund (FWF) Cluster of Excellence QuantA (\href{https://doi.org/10.55776/COE1}{10.55776/COE1}). A. H. and L. L. acknowledge funding from the Austrian Science Fund (FWF) within the DK-ALM (\href{https://doi.org/10.55776/W1259}{10.55776/W1259}). L. L.  acknowledges funding from a joint-project grant from the FWF (No.\,I-4426).

\appendix 
\section{Dynamics of internal states in the presence of imaging light}\label{appA}
\begin{figure}[!ht]
         \includegraphics[width=1.\columnwidth]{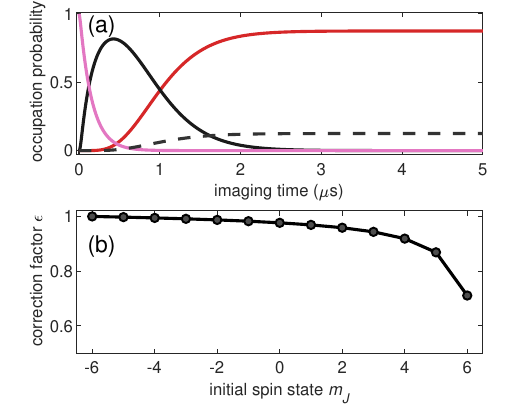}
        \caption{(a) Occupation probabilities in the initial sublevel $\ket{g,0}$ (pink line), all intermediate sublevels from $\ket{g,m_J=-1}$ to $\ket{g,m_J=-5}$ (black line), and the lowest sublevel $\ket{g,m_J=-6}$ (red line) of the ground state, as a function of the exposure time to the imaging light. The dashed line represents the occupation probability in $\ket{e,m_J=-7}$. (b) Relative deviation on the atom number as a function of the imaged spin state for an imaging pulse of \SI{15}{\micro \second}.}
    \label{fig:fig2_supp}
\end{figure}

To quantify the population distribution in the different Zeeman sublevels of the ground state, we perform Stern-Gerlach absorption imaging using a $\sigma^-$-polarized light pulse tuned on resonance with a transition at \SI{401}{\nano \meter} \cite{Frisch2012nlm}. 
This process optically pumps atoms from the excited Zeeman sublevels to $\ket{g,-6}$, altering the sample's effective absorption cross-section. 
Consequently, precise measurement of the spin populations requires to consider the $m_J$-dependence of the average number of photons absorbed by the atoms during the imaging pulse.
To analyze the optical pumping dynamics, we simulate the behavior of atoms in excited ground-state Zeeman sublevels $\ket{g,m_J}$ exposed to the imaging light using the Lindblad master equation with the \textit{Julia} package \emph{QuantumOptics.jl} \cite{kramer2018ajf}, taking into account our experimental parameters.

In Fig.\,\ref{fig:fig2_supp}(a), we show the temporal evolution of atoms initially in $\ket{g,0}$ as they interact with the imaging light. Within \SI{2}{\micro\second}, the majority is transferred to the absolute ground state $\ket{g,-6}$ (red line) and cycles with the Zeeman sublevel $\ket{e,-7}$ of the 401-nm transition (dashed black line). Thus, during our imaging duration of \SI{15}{\micro\second}, the light primarily interacts with the cross-section $\sigma_{-6}$ in $\ket{g,-6}$. From this dynamics, we calculate the number of scattered photons $\nu_{m_J}$ starting from an initial state $\ket{g,m_J}$ during the imaging time and define a correction factor $\epsilon(m_J)=\nu_{m_J}/\nu_{-6}$ that can be used to rescale the atom number.
Figure \ref{fig:fig2_supp}(b) shows the values of $\epsilon$ as a function of the initial state $\ket{g,m_J}$
The largest correction occurs when atoms are initially in $\ket{g,+6}$ ($\epsilon$ = \SI{70}{\percent}). 
For the highest spin state analyzed in this paper $\ket{g,0}$, the correction is close to one ($\epsilon$ = \SI{98}{\percent}).

\section{Feshbach spectrum}\label{appB}

\begin{figure}[b]
         \includegraphics[width=1.\columnwidth]{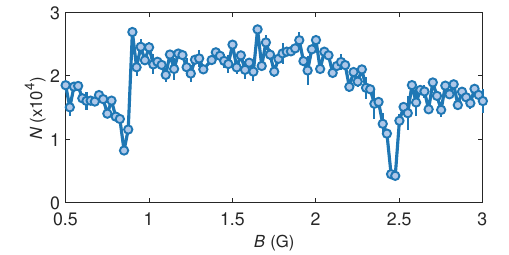}
        \caption{Feshbach spectroscopy for atoms in $\ket{g, -4}$ as a function of the magnetic field $B$. Each data point is the mean of 3 to 5 repetitions and is simply connected with a line.}
    \label{fig:fig3_supp}
\end{figure}

In this work we assume that the dynamics of atom loss and depolarization can be accurately described by considering only two-body relaxation and exchange processes. 
This assumption is valid as long as one- and three-body loss processes can be neglected.
While our one-body loss is negligible on the timescales of the experiment, special care has to be taken to minimize three-body recombination. For this, we perform a high-resolution Feshbach spectroscopy to ensure that the magnetic field is tuned away from any Fano-Feshbach resonances in the spin states of interest.

We prepare a thermal cloud in $\ket{g,-4}$ at \SI{1.9}{G}, then ramp the magnetic field magnitude for 3 ms to a value ranging from \SI{0.5}{} to \SI{3.0}{G}. Subsequently, we hold the atoms in the trap for 20 ms before finally measuring the spin population. Figure \ref{fig:fig3_supp} shows two narrow Feshbach resonances at \SI{0.9}{} and \SI{2.4}{G}. Between \SI{1.1}{} and \SI{2.1}{G}, the atomic population shows a plateau, indicating that the loss processes are independent of the magnetic field magnitude and thus unaffected by the presence of Feshbach resonances. Consequently, at \SI{1.9}{G}, three-body losses can be safely disregarded, when compared to the very fast two-body losses observed in our experiments.

\end{document}